\documentclass[11pt]{article}
\usepackage{jheppub}
\usepackage{amsmath,amssymb,amsfonts,amsthm,mathtools}
\usepackage{revsymb}
\usepackage{bm}
\usepackage[dvipsnames]{xcolor}
\usepackage{hyperref}
\usepackage[utf8]{inputenc}
\usepackage[titletoc]{appendix}
\usepackage{subfigure}

\makeatletter
\def\@fpheader{\relax}
\makeatother

\numberwithin{equation}{section}
\allowdisplaybreaks

\newcommand{\bra}[1]{\langle #1 |}
\newcommand{\ket}[1]{| #1 \rangle}

\newcommand\be{\begin{equation}}
\newcommand\ee{\end{equation}}
\newcommand\bea{\begin{eqnarray}}
\newcommand\eea{\end{eqnarray}}
\newcommand\ba{\begin{array}}
\newcommand\ea{\end{array}}

\newcommand\eref[1]{(\ref{#1})}
\newcommand\bc{\begin{center}}
\newcommand\ec{\end{center}}

\newcommand\comment[1]{}

\DeclareRobustCommand{\w}{\bm{w}}
\title{First Principles Phenomenology of $\bm{H_0}$}

\author[a]{Vishnu Jejjala,}
\author[b]{Michael Kavic,}
\author[c]{Djordje Minic,}
\author[c]{Tatsu Takeuchi\,}

\affiliation[a]{Mandelstam Institute for Theoretical Physics, School of Physics, NITheCS, and CoE-MaSS,\\
University of the Witwatersrand, Johannesburg, WITS 2050, South Africa}
\affiliation[b]{Department of Chemistry and Physics, SUNY Old Westbury, Old Westbury, NY 11568, U.S.A.}
\affiliation[c]{Department of Physics, Virginia Tech, Blacksburg, VA 24061, U.S.A.}

\emailAdd{vishnu@neo.phys.wits.ac.za}
\emailAdd{kavicm@oldwestbury.edu}
\emailAdd{dminic@vt.edu}
\emailAdd{takeuchi@vt.edu}

\abstract{
In this letter we discuss infinite statistics and motivate its r\^ole in quantum gravity.
Then, we connect infinite statistics to a dynamical form of dark energy, and we obtain an expression for the evolution of the Hubble parameter that we compare to observation.
The equation of state parameter $w_\text{eff} < -1$ in this framework.
}


\begin{document}

\maketitle
\parskip=.5\baselineskip

\section{Introduction}\label{sec:intro}

The Universe is expanding~\cite{1922ZPhy,1927ASSBL,hubble1929relation}.
Moreover, this expansion is accelerating~\cite{Riess:1998cb,Perlmutter:1998np}.
Dark energy supplies the engine for the current epoch of accelerated expansion that commenced when the Universe was approximately $9.8$~Gyr old.
The Hubble constant $H_0$ gives the rate at which the scale factor $a(z)$ in a Friedmann--Robertson--Walker (FRW) Universe is changing today ($z=0$).
Determination of $H_0$ from the cosmic microwave background (CMB), an early-time observable ($z\sim 1000$), yields $H_0 = 67.36 \pm 0.54\;\mathrm{km\;s^{-1} Mpc^{-1}}$~\cite{Aghanim:2018eyx}, whereas measurements based on late-time observables ($z \alt 10$), for example, on type Ia supernov\ae, give the value of $H_0 = 73.3 \pm 0.8\,\mathrm{km\;s^{-1} Mpc^{-1}}$~\cite{Verde:2019ivm}.
Because the disagreement between high-redshift and low-redshift measurements is at the four to six $\sigma$ level, we must confront this Hubble tension from the perspective of a fundamental theory~\cite{DiValentino:2021izs}.

To resolve the discrepancy, one can easily imagine that the cosmological constant, $\Lambda$, which computes the vacuum energy density, is not constant at all but changes with time.
In other words, the dark energy is dynamical.
We aim to determine the dynamics of dark energy in as model independent a fashion as possible.
We presuppose a consistent theory of quantum gravity such as string theory, but we are largely agnostic about its details.

The argument for having a dynamical dark energy arises from two basic assumptions: 
(1) the fundamental non-locality of the effective spacetime description, and (2) Lorentz covariance.
These assumptions are ultimately conservative.
The former is an intrinsic feature of low-energy physics whenever a fundamental length scale is present.
The simplest example is the lowest Landau level of a two dimensional electron gas in the presence of a perpendicular magnetic field in the $\hat{z}$-direction.
Upon taking the electron mass to zero, the momentum $p_x$ becomes proportional to the $y$-coordinate, and the canonical commutation relation between position and momentum operators yields a non-commutative geometry~\cite{landau1958quantum,Fradkin:2013sab}.
This means there is non-locality as there is a fundamental length scale in the infrared.
Such a fundamental length scale is intrinsic in Lorentz covariant models of quantum gravity, which also incorporate a consistent coupling of Standard Model-like matter degrees of freedom and gravity, such as string theory~\cite{Green:1987sp}.
The presence of a fundamental length leads naturally to spacetime non-commutativity~\cite{Snyder:1946qz, Connes:1997cr, Seiberg:1999vs}.
(For a more detailed discussion of the r\^ole of intrinsic non-commutativity in string theory and quantum gravity in general, see also~\cite{Freidel:2017wst, Freidel:2017nhg, Berglund:2020qcu} and references in the review~\cite{Minic:2020oho}.)
The essential ingredient of Lorentz covariance is built into general relativity from which the Friedmann equations for the scale factor $a$ are derived upon employing the additional premise, corroborated by observation, that the Universe is homogeneous and isotropic on large scales with the curvature of spatial sections being small.

The consequence of these basic assumptions is that the partonic non-local degrees of freedom of a Lorentz-covariant formulation of quantum gravity with an intrinsic fundamental length obey infinite statistics~\cite{Doplicher:1971wk, Doplicher:1973at, Greenberg:1989ty, Strominger:1993si}.
This is the only statistics consistent with both (1) non-locality and (2) Lorentz covariance.
From this statistical principle, we compute the density of states, which then allows us to deduce the functional form of $\Lambda(z)$.
Feeding this in to the Friedmann equations, we may calculate the Hubble parameter $H(z) = \dot{a}/a$ in terms of the redshift $z$~\cite{Jejjala:2020lhg}.
The functional form of $H(z)$ is compared to astronomical measurements of $H(z)$ at various redshifts $z$, and we find phenomenological agreement to within $5.8$\% as well as a value of $H_0 = 68.5\; \mathrm{km\;s^{-1}\,Mpc^{-1}}$.
However, we find that the effective equation of state parameter for dark energy is necessarily bounded above so that $w_\text{eff} \le -1$.
Observations place experimental constraints on how negative $w_\text{eff}$ can be.

The organization of this letter is as follows.
In Section~\ref{sec:infinitestatistics}, we review infinite statistics and motivate its r\^ole in quantum gravity.
In Section~\ref{sec:darkenergy}, we connect infinite statistics to the dark energy.
In Section~\ref{sec:H}, we obtain an expression for $H(z)$ that we compare to observation.
In Section~\ref{sec:w}, we compute the equation of state parameter for dark energy.
Finally, in Section~\ref{sec:discussion}, we conclude and offer a prospectus for future work.

\section{Infinite statistics}\label{sec:infinitestatistics}

The algebra of harmonic oscillators can be written
\be
\big[\,\hat{a}_i^{\phantom{\dagger}},\, \hat{a}_j^\dagger\,\big]_q 
\;=\; \hat{a}_i^{\phantom{\dagger}} \hat{a}_j^\dagger - q\,\hat{a}_j^\dagger \hat{a}_i^{\phantom{\dagger}} 
\;=\; \delta_{ij} \;. 
\label{eq:qdef}
\ee
Setting $q=1$ is the bosonic case, and~\eqref{eq:qdef} describes the usual quantum harmonic oscillator.
Similarly, setting $q=-1$ in~\eqref{eq:qdef} is the fermionic case in which the commutator is replaced by an anti-commutator.
These yield, respectively, particles with Bose--Einstein and Fermi--Dirac statistics~\cite{Greenberg:1989ty}.

Setting $q=0$ in~\eqref{eq:qdef} also yields a familiar statistics, \textit{viz.}, infinite statistics, corresponding to the Cuntz oscillators~\cite{voiculescu1992free, Gopakumar:1994iq}.
Explicitly, Cuntz oscillators satisfy the algebra
\be
\hat{a}_i^{\phantom{\dagger}} \ket{0} \;=\; 0 \;, \qquad 
\hat{a}_i^{\phantom{\dagger}} \hat{a}_j^\dagger \;=\; \delta_{ij} \;, \qquad 
\sum_i \hat{a}_i^\dagger \hat{a}_i^{\phantom{\dagger}} \;=\; \mathbf{1} - \ket{0}\bra{0}\;.
\ee
Because there are no further relations, the raising and lowering operators neither commute nor anti-commute.
This means the ordering is important: $\hat{a}_i^\dagger \hat{a}_j^\dagger \ne \pm\, \hat{a}_j^\dagger \hat{a}_i^\dagger$.
Respecting this property, the number operator is given by a sum on words~\cite{Greenberg:1989ty,Halpern:2001hg}:
\be
\hat{N} \;=\; \sum_{k=1}^\infty \sum_{i_1} 
\hat{a}_{i_1}^\dagger \Big( \sum_{i_2} \hat{a}_{i_2}^\dagger \ldots \Big( \sum_{i_k} \hat{a}_{i_k}^\dagger \hat{a}_{i_k}^{\phantom{\dagger}} \Big) \ldots \hat{a}_{i_2}^{\phantom{\dagger}} \Big) \hat{a}_{i_1}^{\phantom{\dagger}} \;. 
\label{eq:numop}
\ee
Even in the single oscillator case, we have
\be
\hat{N}_1 
\;=\; \sum_{k=1}^\infty (\hat{a}^\dagger)^k (\hat{a})^k 
\;=\; \frac{\hat{a}^\dagger \hat{a}}{1-\hat{a}^\dagger \hat{a}} 
\;.
\ee
The summand $(\hat{a}^\dagger)^k (\hat{a})^k$ counts the presence of a particle in any given excitation.
When there are more oscillators, the number operator~\eqref{eq:numop} accounts for distinguishable particles, which then satisfy Boltzmann statistics.

The Cuntz algebra is intrinsic to matrix theories.
If we have $n$ independent random matrices, working in the large-$N$ limit, the Fock space obtained from acting the matrices on the vacuum, \textit{i.e.}, by taking $\hat{M}_1\hat{M}_2\ldots |0\rangle$, is isomorphic to the Cuntz algebra~\cite{Gopakumar:1994iq}.
This operation may, for instance, be natural in the Matrix model for M-theory~\cite{Banks:1996vh}.
In the BFSS Matrix model, gravitons are bound states of D$0$-branes.
The gravitational interaction and the geometry of spacetime arise from off-diagonal matrix elements corresponding to open string degrees of freedom stretching between branes.
The D$0$-branes (the partonic degrees of freedom) obey infinite statistics.

The quantum statistics associated to the Cuntz algebra can also be realized in other macroscopic settings in quantum gravity.
In~\cite{Strominger:1993si}, the scattering of extremal black holes is described in terms of infinite statistics.
In~\cite{Jejjala:2007hh, Jejjala:2007rn}, the authors argue that dark energy quanta arising from the Matrix theory framework obey infinite statistics and follow the Wien distribution.
This latter observation is an impetus for the present work.

\section{Dynamical dark energy}\label{sec:darkenergy}

To derive the behavior of the dark energy density as a function of time, we begin by observing that the path integral should have both an ultraviolet cutoff corresponding to a high energy scale at which new physics enters and an infrared cutoff corresponding to the scale of non-locality.
In anti-de Sitter (AdS) space, this low energy cutoff is determined by the radius of curvature~\cite{Heemskerk:2009pn}.
In an approximately de Sitter (dS) space, the scale is set by the cosmological horizon.
Thus,
\be
E_\text{IR} \;=\; \frac{E_0}{a} \;=\; E_0\, (1+z) \;,
\label{eq:eir}
\ee
where $E_0$ is the infrared cutoff today, at $z=0$.
As usual, the expression $a^{-1} = (1+z)$ relates the scale factor of the FRW cosmology to the redshift.

Infinite statistics in an effective four dimensional background tells us that the dark energy quanta follows (quantum) Boltzmann statistics, which we express in terms of the Wien distribution,
\be
I_\text{DE}(E, E_0) \;=\; A\, E^3\, e^{-B E/E_0} \;, 
\label{eq:ide}
\ee
with $A$ and $B$ being dimensionless constants~\cite{Jejjala:2007hh}.
The observed vacuum energy density arises from integrating the previous expression:
\be
\rho_\text{vac} (E_\text{UV}) 
\;=\; \frac{\Lambda(E_\text{UV})}{8\pi G_\text{N}} 
\;=\; \int_0^{E_\text{UV}} dE\ I_\text{DE}(E, E_0) 
\;.
\label{eq:rhovac}
\ee
(In our conventions, $\Lambda$ has units $[\text{time}]^{-2}$.)
Performing the integral, we find
\be
\rho_\text{vac}(z) 
\;=\; \frac{\Lambda(z)}{8\pi G_\text{N}} 
\;=\; \rho_* \big[\, 1 - b(\xi) \,\big] \;, 
\label{eq:Lofz}
\ee
where
\be
\rho_* \;=\; \frac{6A}{B^4}\, E_0^4 \;, \qquad 
b(\xi) \;=\; \left( 1 + \xi + \frac12 \xi^2 + \frac16 \xi^3 \right) e^{-\xi} \;,
\label{eq:params1}
\ee
and
\be
\xi \;=\; \frac{B E_\text{UV}}{E_0} \;=\; \frac{\xi_0}{1+z} \;, \qquad 
\xi_0 \;=\; \frac{B\mu}{E_0^2} \;,
\label{eq:params2}
\ee
with $\sqrt\mu$ a reference energy scale related to the mixing of ultraviolet and infrared degrees of freedom due to non-locality~\cite{Jejjala:2020lhg}.
The scaling $\rho_* \sim E_0^4$ is analogous to the Stefan--Boltzmann law, which teaches that the power of the radiation emitted by a blackbody goes like $T^4$.
From~\eqref{eq:Lofz}, we see that
\be
\chi \;=\; \frac{\Lambda(z)}{\Lambda(0)} \;=\; \frac{1-b(\xi)}{1-b(\xi_0)} \;. 
\label{eq:chi}
\ee
In the $z\to \infty$ limit (early times), $\xi$ goes to zero.
Thus, $b(\xi) \approx 1$, and the ratio~\eqref{eq:chi} vanishes.
Notice that $B$ and $\mu$ only appear in $\xi_0$, so they cannot be individually determined, but $\sqrt{B\mu}$ sets a scale.
The constant $A$ only appears in $\rho_*$ so it cancels in the ratio.
We observe as well that for $\xi\sim\xi_0 \gg 1$, $b(\xi)$ vanishes.
In this limit, $\Lambda(z) = \mathrm{constant}$.

\section{Evolution of the Hubble parameter}\label{sec:H}

Because the vacuum energy density is a function of redshift, which we use as a proxy for time, the Friedmann equation is modified from its standard form.
We now have
\be
H(z)^2 \;=\; \left( \frac{\dot a}{a} \right)^2 \;=\; 
H_0^2 \Big[\, \Omega_m (1+z)^3 + \Omega_\Lambda(z) \,\Big] \;, 
\label{eq:Hofz}
\ee
where, adopting the conventions of~\cite{Weinberg:2008zzc},
\be
\Omega_m \;=\; \frac{\rho_{m,0}}{\rho_{c,0}} \;, \qquad 
\Omega_\Lambda(z) \;=\; \frac{\rho_\Lambda(z)}{\rho_{c,0}} \;=\; \frac{\Lambda(z)}{3H_0^2} \;, \qquad 
\rho_{c,0} \;=\; \frac{3 H_0^2}{8\pi G_\text{N}} \;.
\label{eq:definitions}
\ee
In writing the expression~\eqref{eq:Hofz}, we have dropped the radiation term since it is negligible and work in a Universe with approximately flat spatial sections.
We may recast~\eqref{eq:Hofz} as
\be
H(z) \;=\; H_0 \bigg[\, \Omega_m (1+z)^3 + \Omega_\Lambda(0) \;\frac{\Lambda(z)}{\Lambda(0)} \,\bigg]^\frac{1}{2} \;, 
\label{eq:HH}
\ee
where we recognize that the second term in square brackets as being proportional to $\chi$ from~\eqref{eq:chi}.
Today,
\be
\Omega_m + \Omega_\Lambda(0) \;=\; 1 \;,
\ee
with $\Omega_m\approx 0.3$ and $ \Omega_\Lambda(0) \approx 0.7$~\cite{Aghanim:2018eyx}.

Ref.~\cite{Magana:2017nfs} has compiled $51$ measurements of the Hubble parameter $H(z)$ at various redshifts in the range $0 < z < 2.4$, collected from~\cite{Stern:2009ep,Zhang:2012mp,Moresco:2012jh,Moresco:2015cya,Moresco:2016mzx,Ratsimbazafy:2017vga,Gaztanaga:2008xz,Blake:2012pj,Chuang:2012qt,Anderson:2013oza,Oka:2013cba,Font-Ribera:2013wce,Delubac:2014aqe,Wang:2016wjr,Alam:2016hwk,Bautista:2017zgn}.
Following the terminology of~\cite{Magana:2017nfs}, $H(z)$ can be computed using the so called ``differential age'' (DA)~\cite{Jimenez:2001gg,Stern:2009ep,Zhang:2012mp,Moresco:2012jh,Moresco:2015cya,Moresco:2016mzx,Ratsimbazafy:2017vga} and ``clustering'' methods~\cite{Gaztanaga:2008xz,Blake:2012pj,Chuang:2012qt,Anderson:2013oza,Oka:2013cba,Font-Ribera:2013wce,Delubac:2014aqe,Wang:2016wjr,Alam:2016hwk,Bautista:2017zgn}.

The DA method~\cite{Jimenez:2001gg} relies on the relation 
\begin{equation}
H(z) \;=\; \dfrac{\dot{a}}{a} \;=\; -\dfrac{1}{1+z}\;\dfrac{dz}{dt}\;.
\end{equation}
To determine $dz/dt$ at a particular $z$, the age difference $\Delta t$ of two passively evolving galaxies within an ensemble with small redshift difference $\Delta z$ is determined from their difference in their metalicities.

The clustering method~\cite{Seo:2003pu,Eisenstein:2005su,Percival:2007yw} relies on looking for the imprint of Baryon Acoustic Oscillations (BAO)~\cite{Peebles:1970ag,Sunyaev:1970eu} in the distribution of galaxies. It uses data from the Sloan Digital Sky Survey (SDSS)~\cite{Gaztanaga:2008xz,Chuang:2012qt,Anderson:2013oza,Oka:2013cba,Font-Ribera:2013wce,Delubac:2014aqe,Wang:2016wjr,Alam:2016hwk,Bautista:2017zgn},
and the WiggleZ Dark Energy Survey~\cite{Blake:2012pj}.

\begin{figure}[h!]
\subfigure[]{\includegraphics[width=7cm]{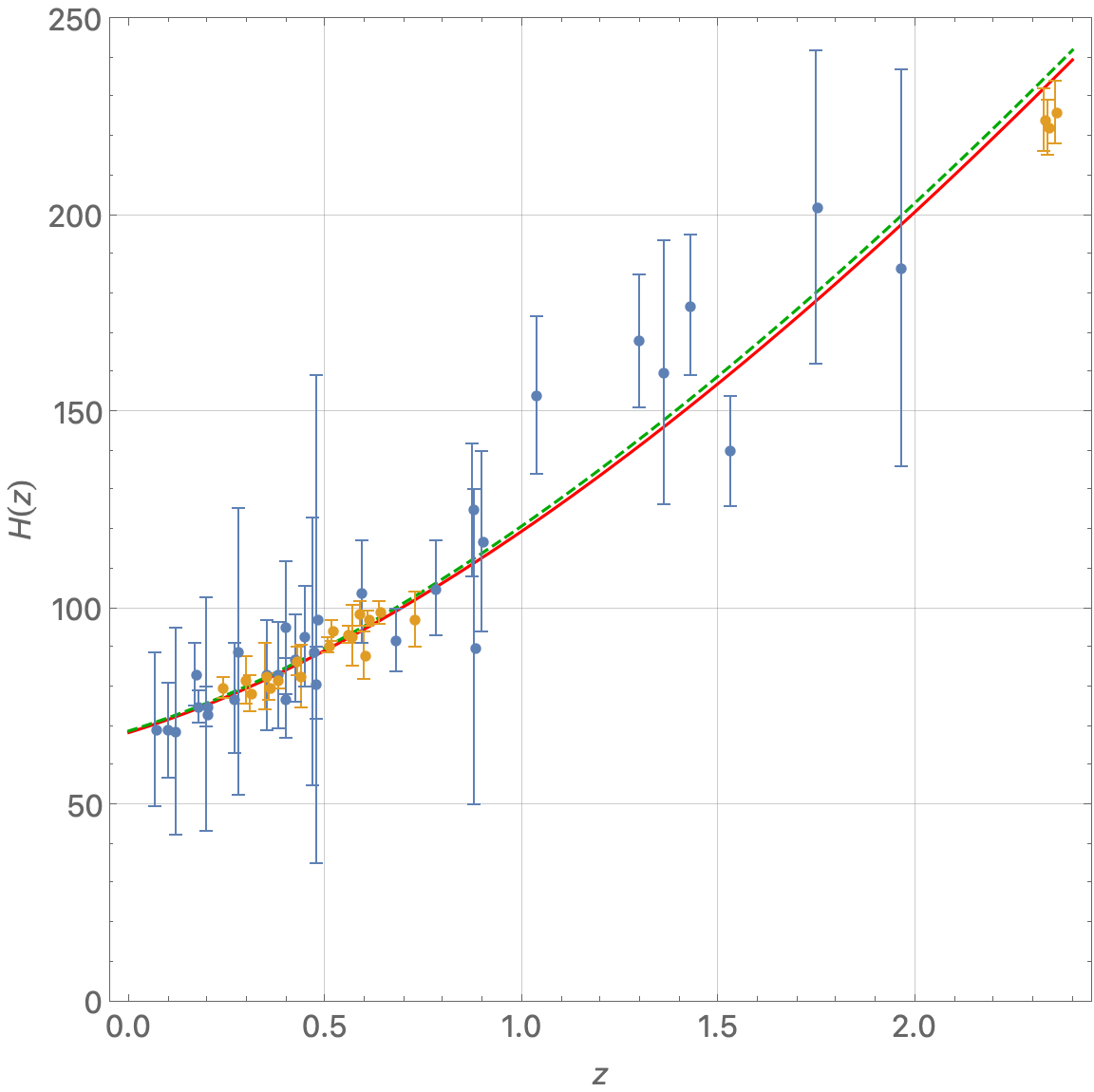}}
\hspace{1cm}
\subfigure[]{\includegraphics[width=7cm]{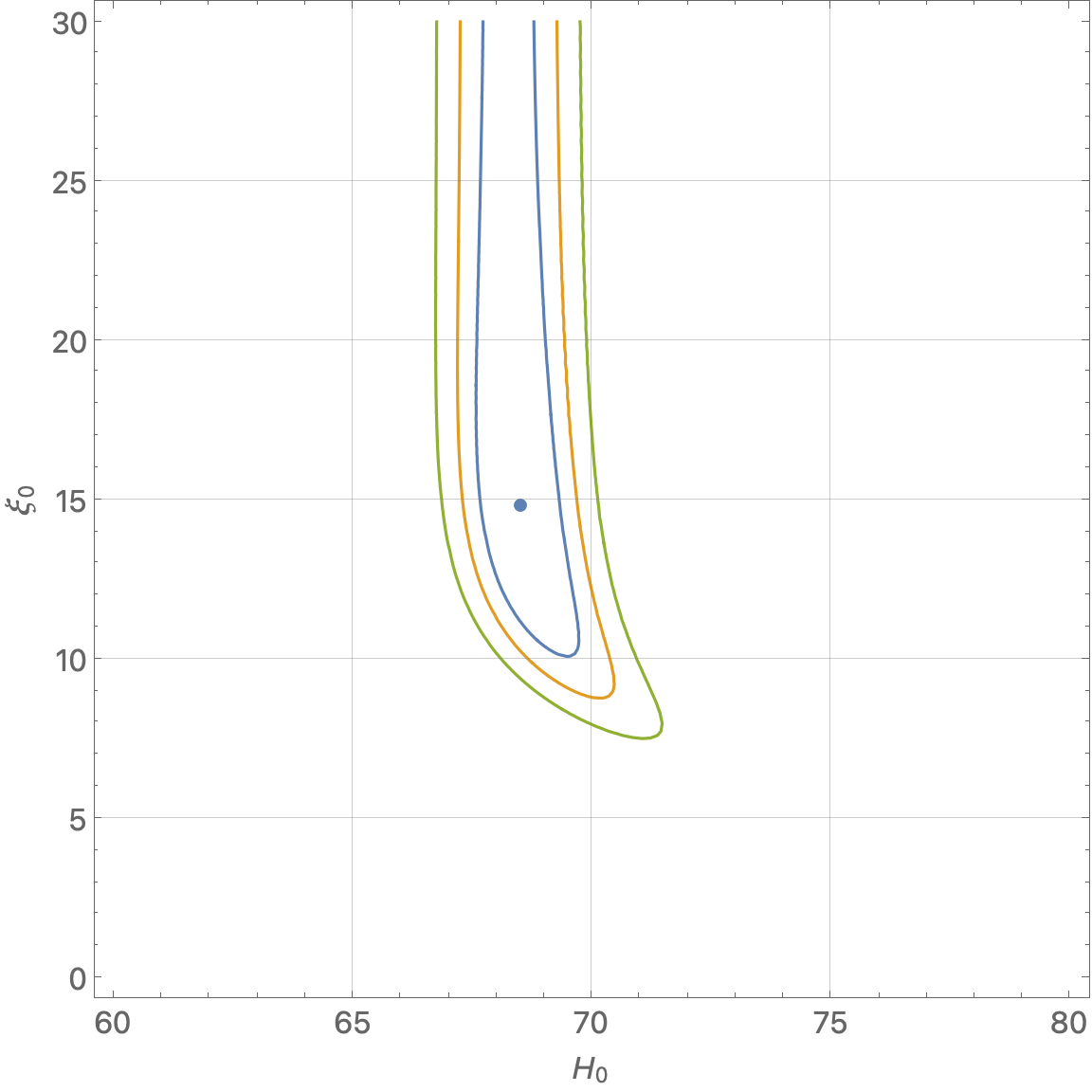}}
\caption{
(a) Red line: best fit of~\eqref{eq:HH} to the $51$ measurements of $H(z)$
listed in Table~1 of~\cite{Magana:2017nfs}.
The green dashed line is an unweighted fit of the observation.
The red line is the $\chi^2$ fit that takes into account the error bars on the data.
The blue dots are ``differential age'' measurements, and the orange dots are ``clustering'' measurements.
(b) The best fit point and the $1$, $2$, and $3\sigma$ likelihood contours.
}
\label{fig:fitTT}
\end{figure}

We perform a least-squares-fit of the function given in~\eqref{eq:HH} to the data tabulated in~\cite{Magana:2017nfs}.
The data points together with the result of our fit are shown in Figure~\ref{fig:fitTT}.
The fit yields:
\be
H_0 \;=\; 68.9\;\mathrm{km\;s^{-1}\,Mpc^{-1}} \;, \qquad 
\xi_0 \;=\; 18.4 \;.
\ee 
If we perform a $\chi^2$-fit using the data with the errors tabulated in~\cite{Magana:2017nfs}, the parameters are slightly different:
\be
H_0 \;=\; 68.5\;\mathrm{km\;s^{-1}\,Mpc^{-1}} \;, \qquad 
\xi_0 \;=\; 14.8 \;.
\ee 
We as well show the best fit contours corresponding to confidences of $1$, $2$, and $3\sigma$.

In the functional form of the fit, only the combination $\xi_0$ appears.
Thus, the precise numerical value of the reference energy $\sqrt{\mu}$ is not especially meaningful.
As a result of this ambiguity, the ratio $E_0/\sqrt{\mu}$ can also be rescaled by tuning the coefficient $B$ in~\eqref{eq:ide}.
It is notable that the fit extracts a value for $H_0$ that agrees more with that determined from the CMB observations ($3\sigma$) rather than that from supernova Ia observations ($5.5\sigma$).
The mean absolute error, computed as
\be
\text{mean absolute error} \;=\; \left\langle \Big| \frac{\text{data point} - \text{fitted value}}{\text{data point}} \Big| \right\rangle \;,
\ee
is $5.8$\%.
Notice that the point $\xi_0\to \infty$ corresponding to a non-dynamical dark energy --- \textit{i.e.}, the usual cosmological constant --- is within the contour.

\section{Equation of state parameter}\label{sec:w}

Let us compute the equation of state parameter for dynamical dark energy.
The calculation we perform is independent of the statistics of the dark energy quanta.
In the end, we specialize to the case of infinite statistics.
For convenience, we introduce a new variable $x=(1+z)$.

From~\eref{eq:definitions}, it follows that
\be
\Omega_\Lambda(x) \;=\; \Omega_\Lambda(0) \frac{\Lambda(x)}{\Lambda(0)} \;,
\ee
where, using~\eref{eq:rhovac},
\be
\Lambda(x) \;=\; 8\pi G_N \int_0^{E_\text{UV}(x)} dE\ \rho_\Lambda(E) \;.
\label{eq:lamx}
\ee
In Sections~\ref{sec:darkenergy} and~\ref{sec:H}, the precise value of the energy scale $\sqrt\mu$ was unimportant. 
Stemming from non-locality, we will now take $\sqrt\mu$ to be an intermediate scale associated to the mixing of degrees of freedom in the ultraviolet and infrared~\cite{Freidel:2018apz,Freidel:2021wpl}:
\be
\mu \;=\; E_\text{IR} E_\text{UV} \;.
\label{eq:geom}
\ee
This is an additional assumption\footnote{
Often, $\mu\sim m_\text{susy}^2$ references the mass scale at which low energy supersymmetry is broken~\cite{Banks:2000fe,Jejjala:2006jf,Heckman:2018mxl}.
If $E_\text{IR}$ corresponds to the current vacuum energy density and $E_\text{UV}$ to the Planck scale, then $\sqrt\mu\simeq 5\ \text{TeV}$~\cite{Jejjala:2020lhg}.}
we have introduced that, together with~\eref{eq:eir}, tells us
\be
E_\text{UV} \;=\; \frac{\mu}{E_\text{IR}} \;=\; \frac{\mu a}{E_0} \;=\; \frac{\mu}{E_0 x} \;.
\label{eq:euv}
\ee
Differentiating~\eref{eq:lamx}, we derive
\be
\frac{d\Lambda(x)}{dx} \;=\; -(8\pi G_N) \rho_\Lambda(E_\text{UV}(x)) \frac{\mu}{E_0 x^2} \;.
\label{eq:dldx}
\ee
It immediately follows that
\be
\frac{d}{dx} \log \Omega_\Lambda(x) \;=\; \frac{d}{dx} \log \Lambda(x) \;=\; -(8\pi G_N) \frac{\rho_\Lambda(E_\text{UV}(x))}{\Lambda(x)} \frac{\mu}{E_0 x^2} \;.
\ee
This enables us to determine the equation of state parameter for dark energy because in the FRW Universe~\cite{Saini:1999ba,Alam:2003fg},
\be
w_\text{eff} \;=\; -1 + \frac{x}{3} \frac{d}{dx} \log \Omega_\Lambda(x) \;.
\label{eq:weffde}
\ee
(For completeness, we sketch a derivation of this expression in Appendix~\ref{app:w}.)
In combination with~\eref{eq:euv} and~\eref{eq:dldx}, we recast~\eref{eq:weffde} as
\be
w_\text{eff} \;=\; -1 - \left[ \frac{8\pi G_N}{3} \frac{\rho_\Lambda(E_\text{UV}(x))}{\Lambda(x)} E_\text{UV} \right] \;.
\label{eq:weffde2}
\ee
Since the term in square brackets is positive, $w_\text{eff} < -1$.
This is reminiscent of~\cite{Das:2005yj,vandeBruck:2019vzd,Agrawal:2019dlm}.

Infinite statistics enables us to write~\eref{eq:weffde2} explicitly.
Making the substitutions given by~\eref{eq:Lofz},~\eref{eq:params1}, and~\eref{eq:params2}, we find
\be
w_\text{eff} \;=\; -1 - \frac{\xi^4 e^{-\xi}}{18[1-b(\xi)]} \;, \qquad \xi \;=\; \frac{\xi_0}{x} \;.
\ee
Indeed, in the limit where $\xi_0\to\infty$, we recover $w_\text{eff} = -1$, corresponding to a cosmological constant.

\section{Discussion}\label{sec:discussion}

Using general precepts of spacetime non-locality and Lorentz covariance in quantum gravity, we are led inexorably to infinite statistics to describe the statistics of distinguishable quanta.
Associating these quanta with dark energy, we propose a density of states that follows the Wien distribution.
Modifying the Friedmann equation to account for the fine structure of dark energy, we then find that the Hubble parameter $H_0$ is a function of the redshift $z$.
This is consistent with observational data that supports a lower value of $H_0$ from high redshift measurements of the early Universe.
Infinite statistics leads us to conclude that the equation of state parameter $w_\text{eff} < -1$.
In particular, this rules out quintessence models of an accelerating Universe with $-1 < w_\text{eff} < -\frac13$.

Suppose we model dark energy with a scalar field.
The equation of state parameter $w_\text{eff} < -1$ can be accommodated by a wrong sign kinetic term.
The scalar field is a ghost or phantom that experiences the opposite spacetime signature to ordinary matter~\cite{Gibbons:2003yj}.
There are typically caustics where the equations of motion break down, but these can be avoided through judicious choices of the Lagrangian~\cite{Gibbons:2003yj}.
Maintaining unitarity comes at the expense of vacuum instability.
If Lorentz invariance is preserved, then there are superluminal modes, and the theory violates classical causality.
(See~\cite{Caldwell:2003vq} and~\cite{Ludwick:2017tox} for a review.)
To evade these pathologies, one can consider a scalar field which rolls up rather than down a potential~\cite{Csaki:2005vq}.
Viable cosmological models with $w_\text{eff} < -1$ are thus not excluded by the data and in fact may provide the best fit to the data (\textit{cf.}\ Table~4 in~\cite{Aghanim:2018eyx}).

The current measurements of $w_\text{eff}$ and their uncertainties as well constrain the fit parameter $\xi_0$.
The value $\xi_0\to\infty$ which corresponds to equating the cosmological constant to the dark energy is allowed by the data and is consistent with the hypothesis of infinite statistics.

The present datasets are incomplete.
Between the high redshift measurements from the cosmic microwave background and low redshift measurements from a variety of observations including type Ia supernov\ae\ are intermediate redshifts that will be probed in the near future, for instance, by the Square Kilometre Array's investigation into the Dark Ages when the first stars and galaxies formed.
Our framework predicts the values for $H_0$ in this intermediate regime.
Future observations of Cepheid variables by, for instance, the James Webb Space Telescope, will enable improved measurements of the Hubble parameter at low redshift~\cite{Freedman:2021ahq}.
We anticipate that these measurements will help resolve the Hubble tension.

\acknowledgments

We thank P.~Berglund, E.~Bianchi, D.~Edmonds, L.~Freidel, S.~Horiuchi, T.~H\"ubsch, J.~Kowalski-Glikman, and R.~G.~Leigh for helpful discussions. 
VJ is supported in part by the South African Research Chairs Initiative of the National Research Foundation, grant number 78554 and the Simons Foundation Mathematical and Physical Sciences Targeted Grants to Institutes, Award ID:509116.
DM and TT are supported in part by the U.S.\ Department of Energy (DE-SC0020262, Task C) and the Julian Schwinger Foundation.

\appendix

\section{Derivation of $\w_\text{eff}$}\label{app:w}

The Friedmann equation tells us that
\be
H(x)^2 \;=\; \frac{8\pi G_N}{3} \sum_i \rho_i(x) \;\approx\; \frac{8\pi G_N}{3} \big( \rho_m(x) + \rho_\Lambda(x) \big) \;.
\ee
Rearranging terms, we have
\be
\rho_\Lambda(x) \;=\; \frac{3H(x)^2}{8\pi G_N} \big( 1 - \Omega_m(x) \big) \;.
\ee
Next, the deceleration parameter is
\be
q(x) \;=\; -\frac{\ddot{a}}{a H(x)^2} \;=\; \frac{4\pi G_N}{3 H(x)^2} \sum_i \big[ \rho_i(x) + 3 p_i(x) \big] \;\approx\; \frac12 + \frac{4\pi G_N}{H(x)^2} p_\Lambda(x) \;.
\ee
Thus,
\be
p_\Lambda(x) \;=\; \frac{H(x)^2}{4\pi G_N} \Big( q(x) - \frac12 \Big) \;.
\ee
The ratio of pressure to energy density gives the effective equation of state parameter for dark energy:
\be
w_\text{eff} \;=\; \frac{p_\Lambda(x)}{\rho_\Lambda(x)} \;=\; \frac{2 q(x) - 1}{3(1-\Omega_m(x))} \;.
\label{eq:poverrho}
\ee

Recall that $x = (1+z) = a^{-1}$.
To simplify the denominator of~\eref{eq:poverrho}, we first note that
\be
\Omega_m(x) \;=\; \left( \frac{H_0}{H(x)} \right)^2 \Omega_{m,0} x^3 \;.
\ee
To obtain an expression for the numerator of~\eref{eq:poverrho}, we then observe that
\be
\frac{dH}{dt} \;=\; \frac{\ddot{a}}{a} - \left( \frac{\dot{a}}{a} \right)^2 \;=\; -H(x)^2 \big( q(x) + 1 \big) \;.
\ee
This means
\be
q(x) \;=\; -\frac{1}{H(x)^2} \frac{dH(x)}{dt} - 1 \;=\; \frac{dx}{dt} \frac{d}{dx} \left[ \frac{1}{H(x)} \right] - 1 \;,
\ee
where
\be
\frac{dx}{dt} \;=\; -\frac{\dot{a}}{a^2} \;=\; -x H(x) \;.
\ee
Since
\be
H(x) \;=\; H_0 \sqrt{ \Omega_{m,0} x^3 + \Omega_\Lambda(x) } \;, 
\ee
we deduce after some straightforward algebra that
\be
2q(x) - 1 \;=\; \frac{3H_0^2}{H(x)^2} \left( \frac{x}{3} \frac{d\Omega_\Lambda(x)}{dx} - \Omega_\Lambda(x) \right) \;.
\ee
Putting all the pieces together,
\be
w_\text{eff}(x) \;=\; \frac{\dfrac{x}{3} \dfrac{d\Omega_\Lambda(x)}{dx} - \Omega_\Lambda(x)}{\Omega_\Lambda(x)}
\;=\; -1 + \frac{x}{3}\frac{d}{dx} \log \Omega_\Lambda(x) \;.
\ee
If $\Lambda(x) = \text{constant}$, we verify that $w_\text{eff}=-1$ as required.

\bibliographystyle{JHEP}
\bibliography{refs}

\end{document}